\documentclass[reprint,twocolumn,showpacs,superscriptaddress,longbibliography,email,floatfix,aps,prl]{revtex4-2}

\usepackage{header,mathrsfs}

\def\Id{\mathbb{1}}

\definecolor{MDG}{rgb}{0,0.55,0.05} 
\definecolor{eggplant}{RGB}{126,93,181}

\begin{document}
\title{Large deviations and conditioned monitored quantum systems: a tensor network approach}

\author{María Cea}
\affiliation{Max-Plank-Institut f\"ur Quantenoptik, Hans-Kopfermann-Str. 1, D-85748 Garching, Germany}
\affiliation{Munich Center for Quantum Science and Technology (MCQST), Schellingstr. 4, D-80799 M\"unchen, Germany}
\author{Marcel Cech}
\affiliation{Institut f\"ur Theoretische Physik and Center for Integrated Quantum Science and Technology, Universit\"at T\"ubingen, Auf der Morgenstelle 14, 72076 T\"ubingen, Germany}
\author{Federico Carollo}
\affiliation{Dipartimento di Fisica, Sapienza Università di Roma, Piazzale Aldo Moro 2, 00185 Rome, Italy}
\affiliation{Centre for Fluid and Complex Systems, Coventry University, Coventry, CV1 2TT, United Kingdom}
\author{Igor Lesanovsky}
\affiliation{Institut f\"ur Theoretische Physik and Center for Integrated Quantum Science and Technology, Universit\"at T\"ubingen, Auf der Morgenstelle 14, 72076 T\"ubingen, Germany}
\affiliation{School of Physics and Astronomy and Centre for the Mathematics and Theoretical Physics of Quantum Non-Equilibrium Systems, The University of Nottingham, Nottingham, NG7 2RD, United Kingdom}
\author{Mari Carmen Bañuls}
\affiliation{Max-Plank-Institut f\"ur Quantenoptik, Hans-Kopfermann-Str. 1, D-85748 Garching, Germany}
\affiliation{Munich Center for Quantum Science and Technology (MCQST), Schellingstr. 4, D-80799 M\"unchen, Germany}

\begin{abstract}
Coexistence of different dynamical phases is a hallmark of glassy dynamics. This is well-studied in classical systems where the underlying theoretical framework is that of large deviation theory. The presence of a similar phase coexistence has been suggested in monitored quantum many-body systems, but the lack of suitable methods has yet prevented a systematic large deviation analysis. Here we present a tensor network framework that allows the application of large deviation theory to large quantum systems.
Building on this, we locate a series of first-order dynamical phase transitions in a monitored discrete-time many-body quantum dynamics, at the level of the trajectory space.
Crucially, our approach provides access not only to large-deviation statistics but also to conditioned quantum many-body states, enabling a microscopic characterization of the dynamical phases and their coexistence. 
\end{abstract}

\maketitle

\textbf{Introduction. ---}
The rapid development of quantum technologies enables
the investigation of a variety of dynamical scenarios where thermalization is suppressed or especially slow~\cite{DAlessio2016}, such as those presenting quantum many-body scars~\cite{bernien_probing_2017,turner_quantum_2018,serbyn_quantum_2021}, constrained dynamics~\cite{olmos2014out,ostmann2019localization,scherg2021observing} or Hilbert-space fragmentation~\cite{Sala_2020_hsf,moudgalya_quantum_2022}. Furthermore, the possibility of monitoring or measuring evolving quantum systems leads to new out-of-equilibrium paradigms, including measurement induced phase transitions~\cite{Li2018,Skinner2019,LiChenFisher2019}, learnability transitions~\cite{barratt_2022_learn,ippolity_2024_learn} or conditioned ensembles~\cite{nahum_2025_bayesian,gopalakrishnan_2026_hydro}. In these settings, the measurement outcomes define ensembles of quantum trajectories through which the dynamics can be scrutinized~\cite{Daley2014,garrahan_thermodynamics_2010}. Such trajectory ensembles play a fundamental role in the study of classical kinetically constrained models (KCMs), introduced to describe slow relaxation in glasses~\cite{FredricksonAndersen1984,ritort_glassy_2003,garrahan_dynamical_2007}. Within this framework, large deviation (LD) methods~\cite{touchette_large_2009,garrahan_thermodynamics_2010} shed light on the coexistence of active and inactive trajectories, i.e., histories with many or few occurrences of a given outcome. In the thermodynamic limit, this coexistence corresponds to a first-order dynamical phase transition in trajectory space, which has been proposed as the defining dynamical signature of glassy behavior~\cite{hedges_dynamic_2009,garrahan_first-order_2009,chandler_dynamics_2010}. In many-body systems, however, accessing LD statistics becomes challenging due to the exponential growth of the trajectory space dimension with both system size and observation time.

Tensor network (TN) tools~\cite{schollwock_density-matrix_2011,orus_practical_2014,banuls_tensor_2023} arise as natural candidates to investigate such dynamical settings, and have been successfully applied to KCMs~\cite{Gorissen2009dmrg,banuls2019using}. In many classical KCMs, a similarity transformation maps the tilted generator onto an effective quantum Hamiltonian~\cite{garrahan_first-order_2009} whose ground state energy yields the scaled cumulant generating function (SCGF) associated with fluctuations of a time-integrated trajectory observable. This allows an almost direct application of standard TN methods to extract LD functions~\cite{banuls2019using,causer2021optimal,causer2023optimal}. In contrast, for trajectory observables in monitored quantum systems, the tilted operator whose dominant eigenvalue yields the SCGF is a non-Hermitian quantum channel acting in Liouville space, and, unlike in many classical KCMs where such mappings exist, a mapping to a local Hermitian Hamiltonian is generally not available. As a consequence, standard variational TN methods cannot be straightforwardly applied. Existing TN methods for quantum dynamics can still be used to simulate individual trajectories of large systems~\cite{Daley2014,cech_2025_revealing}, but they do not provide access to the full trajectory ensemble and its LD properties, crucial for accurate determination of dynamical phases. 

In this work, we address this limitation by developing a TN framework that enables the LD analysis of monitored quantum many-body systems with discrete-time dynamics. The method provides direct access to the dominant eigenvalue and eigenvector of the tilted quantum channel, and thus to the SCGF and the dynamical phase diagram. Remarkably, the formalism provides also access to the ensemble of quantum trajectories conditioned by the measurement outcomes, which uncovers the microscopic nature of the quantum dynamics connected to the dynamical heterogeneity in space-time. As such, our method naturally extends the LD formalism to a broad class of problems studied in the context of learnability transitions or conditioned ensembles.

\begin{figure}[t]
    \centering
     \includegraphics[width=\columnwidth]{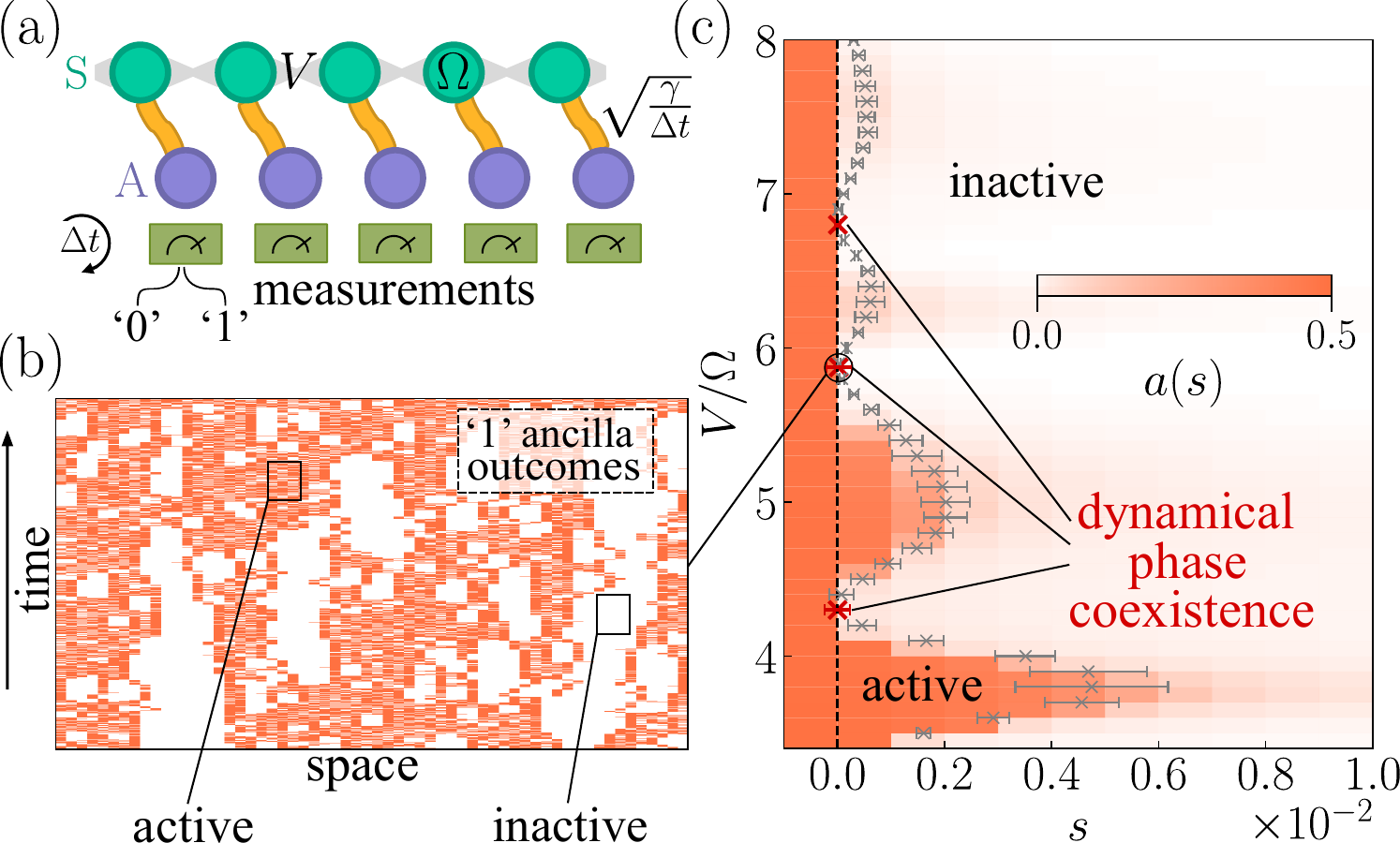}
     \vspace{-20pt}
    \caption{\textbf{Collision model and dynamical phase diagram.} (a) A chain of two-level systems ($\mathrm{S}$, green) interacting according to Eq.~\eqref{eq:Rydberg_Hamiltonian} is coupled site-wise to ancilla qubits ($\mathrm{A}$, purple). At the end of each discrete time interval $\Delta t$, the ancillas are measured and reset. The measurement outcomes define space–time quantum trajectories. (b) Example of a single trajectory for a chain of length $L=60$ at interaction strength $V/\Omega=5.875$. Each pixel represents the measurement outcome for one site at a particular time. (c) Dynamical phase diagram in the interaction-bias $(V/\Omega,s)$ plane obtained from the activity $a(s)$ (see main text) for $L=60$. The dashed line at $s=0$ highlights the physical dynamics. Crosses mark the estimated location of the first-order transition line, with red crosses indicating points where the transition occurs exactly at $s=0$.
    }
    \label{fig:phase_diagram}
\end{figure}

As a concrete application, we consider the collision model of monitored many-body dynamics introduced in~\cite{ciccarello_quantum_2022,Cech_2025}, where a quantum system interacts sequentially with ancillas that are measured and reset at each collision step [cf.~Fig.~\ref{fig:phase_diagram}(a)]. The measurement outcomes reveal complex features~[Fig.~\ref{fig:phase_diagram}(b)], such as dynamical coexistence of active and inactive space-time regions. Applying this framework to a driven-dissipative quantum system motivated by Rydberg atom quantum simulators~\cite{saffman_quantum_2010,labuhn2016tunable,bernien_probing_2017,browaeys2020manybody,ebadi2021quantum}, we systematically identify a series of first-order dynamical phase transition points [Fig.~\ref{fig:phase_diagram}(c)] that are hallmark signatures of glassiness.

\textbf{Collision model setup. ---} 
We consider a system of $L$ qubits ($\mathrm{S}$),  interacting during time intervals of duration $\Delta t$ with a set of $L$ ancilla qubits ($\mathrm{A}$) prepared in the state $\ket{\mathbf{0}_\mathrm{A}} = \bigotimes_{i=1}^L \ket{0_\mathrm{A}}_i$ [cf.~Fig.~\ref{fig:phase_diagram}(a)]. This unitary evolution $U_\mathrm{CM}$ is followed by a projective measurement of the ancillas in the computational basis and their subsequent reset. The resulting stochastic evolution of the system is described by Kraus operators $K_\mathbf{k} =\bra{\mathbf{k}_\mathrm{A}}U_{\mathrm{CM}}\ket{\mathbf{0}_\mathrm{A}}$, where the binary string $\mathbf{k}$ collects the ancilla measurement outcomes~\footnote{For simplicity, we drop the subscript $\mathrm{A}$ in $\mathbf{k}_\mathrm{A}$ from now on.}. Furthermore, the effective open dynamics of the system is described by the quantum channel~\cite{nielsen_quantum_2010} 
\begin{align}
    \mathcal{E}(\rho) = \sum_{\mathbf{k}} K_{\mathbf{k}} \rho K_{\mathbf{k}}^\dagger \, .
    \label{eq:quantum_channel}
\end{align}
For concreteness, we consider that the unitary step is generated by the following collision model Hamiltonian
\begin{align}
    H_\mathrm{CM} = H_\mathrm{S} \otimes \Id + \sqrt{\frac{\gamma}{\Delta t}} \sum_{i = 1}^L (1-n_i) \otimes \tau_i^x \, .
    \label{eq:collision_hamiltonian}
\end{align}
The system Hamiltonian reads
\begin{align}
    H_\mathrm{S} = \Omega \sum_{i=1}^L \sigma_i^x + V \sum_{i=1}^{L-1} n_i n_{i+1} \, ,
    \label{eq:Rydberg_Hamiltonian}
\end{align}
where $\sigma_i^x$ and $\tau_i^x$ denote Pauli-$x$ operators acting on the system and ancilla qubits, respectively, and $n_i = (1-\sigma_i^z)/2$ is the local excitation number. When the interaction strength $V$ is much larger than the Rabi frequency $\Omega$, the system Hamiltonian becomes that of the PXP model~\cite{lesanovsky_interacting_2012,turner_quantum_2018,moudgalya_quantum_2022}.
The second term in Eq.~\eqref{eq:collision_hamiltonian} describes the coupling between system and ancilla qubits. In the continuous-time limit ($\Delta t\to0$), the dynamics reduces to local measurement-induced dephasing at rate $\gamma$~\cite{Cech_2025}. More generally, the ancilla outcomes weakly monitor the local excitation density $\expval{n_i(t)}$, generating space–time trajectories that record the stochastic system dynamics. Even though the stationary state of the system is always the fully mixed one, $\Id/2^L$, in certain parameter regimes the trajectories may show a rich and highly correlated structure [cf.~Fig.~\ref{fig:phase_diagram}(b)].

\begin{figure*}[ht]
    \centering
    \includegraphics[width=\textwidth]{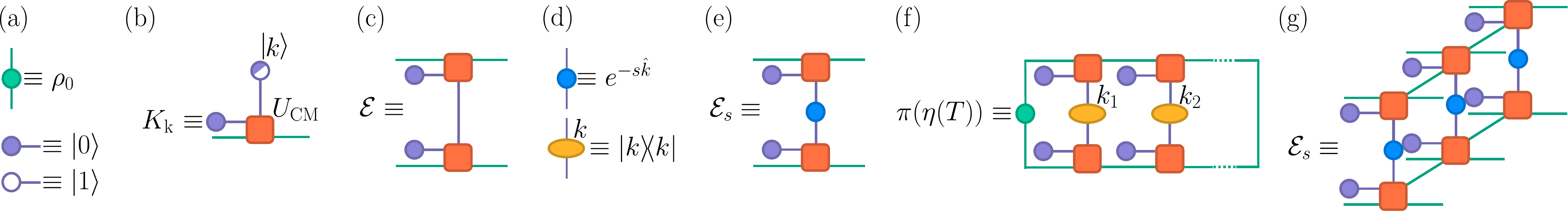}
    \vspace{-20pt}
    \caption{\textbf{
    Tensor network representation of trajectory ensembles.} Tensors are represented as boxes or circles with green (purple) legs corresponding to indices in the system (ancilla) basis. (a) Initial states of the system and ancilla. (b) $k-$th Kraus operator. (c) Quantum channel obtained by contracting the Kraus operators over the ancilla index $k$, interpreted as an MPS transfer matrix. (d) Local bias and projector onto the ancilla state $\ket{k}$ acting on the ancilla legs. (e) Tilted channel obtained by inserting the bias operator on the ancilla legs. (f) MPS contraction representing the trajectory probability $\pi(k_1,\ldots k_T)$. (g) Extension to a many-body chain by approximating the Kraus map as an MPO. 
}
    \label{fig:TN_diagram}
\end{figure*} 

\textbf{Ensemble of trajectories and large deviations. ---}
The set of ancilla measurement outcomes over $T$ collision steps defines a statistical ensemble of trajectories~\cite{touchette_large_2009}, each of them corresponding to a possible space-time record $\eta(T)=\left[\mathbf{k} (t)\right]_{t=1}^{T}$, which occurs with probability $\pi(\eta(T))$. The total number of measurement outcomes equal to 1 is named dynamical activity of the trajectory $\mathcal{A}(\eta(T))$. We define the mean activity $a = \mathcal{A}/(LT)$, and call trajectories with high (low) value of $a$ active (inactive). In the long-time limit, the probability distribution of activity values $P_T(\mathcal{A})$ satisfies a LD principle of the form $P_T(\mathcal{A})\asymp e^{-LT \phi(a)}$, where $\phi(a)$ is the rate function~\cite{touchette_large_2009,garrahan_thermodynamics_2010}.

To probe fluctuations of the dynamical activity it is convenient to introduce a biased ensemble in which trajectories are assigned a weighted probability $e^{-s\mathcal{A}(\eta(T))} \pi(\eta(T))$, for given counting field $s$ that biases the activity. At $s=0$ trajectories are sampled according to $\pi(\eta(T))$, corresponding to the original dynamics. Tuning $s$ away from zero bias the ensemble towards trajectories of atypically high or low activity, allowing one to probe collective behavior in trajectory space and access rare dynamical fluctuations~\cite{garrahan_dynamical_2007,hedges_dynamic_2009,garrahan_thermodynamics_2010}. The partition function of the biased ensemble satisfies also a LD form $Z_T(s) = \langle e^{-s\mathcal{A}} \rangle \asymp e^{LT \theta(s)}$~\cite{touchette_large_2009,garrahan_thermodynamics_2010}, where $\theta(s)$ is called the scaled cumulant generating function. In direct analogy with singularities of the free energy in equilibrium statistical mechanics, non-analyticities of $\theta(s)$ signal phase transitions between distinct dynamical regimes~\cite{lecomte_thermodynamic_2007,garrahan_thermodynamics_2010,goldenfeld2018lectures}. Note also that the functions $\theta(s)$ and $\phi(a)$ are related via a Legendre–Fenchel transform $\phi(a) = \sup_s \left[ -s a - \theta(s) \right]$~\cite{touchette_large_2009}. 

To compute $\theta(s)$ we introduce a tilted quantum channel
\begin{align}
\mathcal{E}_s(\rho) = \sum_{\mathbf{k}} e^{-s \mathcal{A}(\mathbf{k})} K_{\mathbf{k}} \rho K_{\mathbf{k}}^\dagger \, ,
\label{eq:biased_map}
\end{align}
where $\mathcal{A}(\mathbf{k})$ denotes the contribution of the outcome string $\mathbf{k}$ to the total activity. Noticing that $Z_T(s)=\mathrm{Tr}\left[ \mathcal{E}_s^T(\rho_0)\right]$, the SCGF $\theta(s)$ can be obtained from the dominant eigenvalue of $\mathcal{E}_s$. From $\theta(s)$ we can compute the mean activity in the biased ensemble, $a(s) \equiv \langle \mathcal{A} \rangle_s/(LT) = - d\theta(s)/ds$, where $\langle \cdot \rangle_s$ denotes the average over trajectories biased by the field $s$. 

\begin{figure*}[ht]
    \centering
    \includegraphics[width=\textwidth]{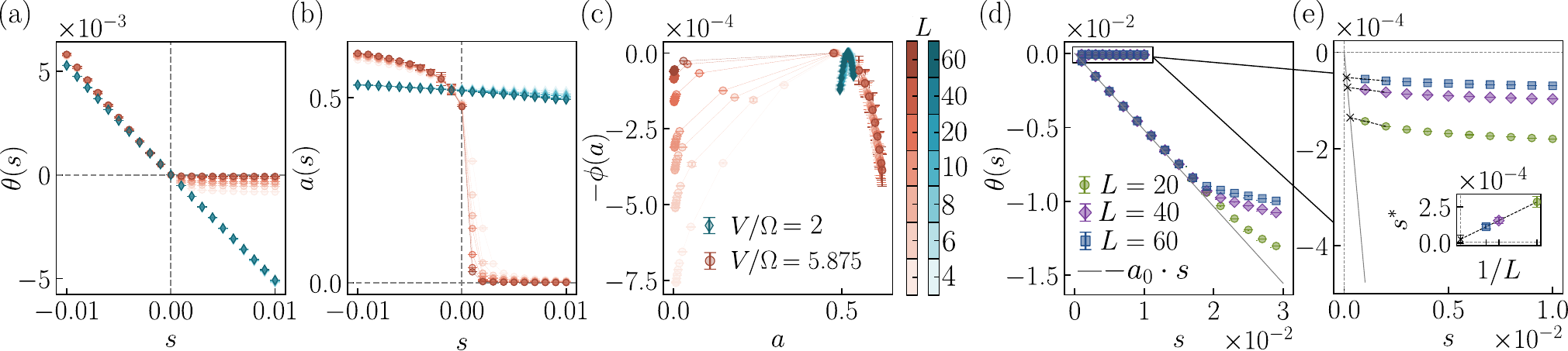}
    \vspace{-20pt}
    \caption{
    \textbf{Signatures of dynamical phase coexistence.}
    (a,b,c): For $V/\Omega =5.875$ (red circles) the SCGF $\theta(s)$ (a) develops a non-analyticity near $s=0$ as system size increases, as the activity $a(s)$ (b) exhibits a rapidly sharpening crossover. The rate function $-\phi(a)$ (c) broadens and exhibits a Maxwell construction, corresponding to the convex hull of an underlying bimodal distribution. In contrast, for $V/\Omega =2$ (blue diamonds), $\theta(s)$ and $a(s)$ are smooth, and $-\phi(a)$ remains narrow with a peak at the stationary value $a\approx0.5$, with no noticeable system size dependence. (d,e): Finite-size analysis of the crossing point of $\theta(s)$ from system sizes $L=20,$ 40, 60. At $s\approx 0$, perturbation theory predicts a linear behavior for the active branch (solid grey line). The crossing is identified by extrapolating the inactive branch until it intercepts this linear behavior. For $V/\Omega =2$ (d) the crossing is far from $s=0$. For $V/\Omega =5.875$ (e) we linearly fit the smallest computed $s>0$ values (dotted black lines) to find the interception point (black crosses). Extrapolating these values with $1/L$ (inset) yields the value $s^* = (2 \pm 3) \times 10^{-5}$, consistent with a transition at $s=0$ in the thermodynamic limit. Results were computed with Trotter step $\Delta t/10$, and bond dimension $D_{\max}=96$, and error bars denote the difference with respect to the results with $D_\mathrm{max}=64$ (see SM for details~\cite{SM}).
    }
    \label{fig:numerics}
\end{figure*}

\textbf{TN representation of trajectory ensembles. ---}
The tilted map $\mathcal{E}_s$ acts as a superoperator in Liouville space whose dimension grows exponentially with the number of sites $L$, rendering a direct numerical treatment infeasible for large many-body systems. However, TNs provide a natural representation for the quantum channel and trajectory ensembles. Figure~\ref{fig:TN_diagram} illustrates it for the case of a single system using the standard TN graphical representation~\cite{schollwock_density-matrix_2011,orus_practical_2014}, in which tensors are depicted as geometric figures with as many legs as indices, and contractions by joined legs. The set of Kraus maps $\{K_k\}$ can be understood as a single rank-3 tensor $K$, where one leg (purple) takes values over the Kraus index $k$ (corresponding to the measured ancilla state) and the other two (green) have the dimension of the system. The individual Kraus operator $K_k$ corresponds to selecting the value $k$ for the ancilla index, and is equivalent to projecting the ancilla onto the state $\ket{k}$ after the collision step [Fig.~\ref{fig:TN_diagram}(b)].

The quantum channel~\eqref{eq:quantum_channel} is obtained by averaging over measurement outcomes, i.e., summing over the $k$ index, and has the structure of an MPS transfer matrix [Fig.~\ref{fig:TN_diagram}(c)], where the bond dimension corresponds to that of the physical system. The initial state plays the role of a boundary condition and applying the channel to it generates the averaged evolved state of the system on the virtual space (green legs). In particular, the stationary state is obtained in the limit of infinitely many applications. To obtain the tilted channel $\mathcal{E}_s$ defined in~\eqref{eq:biased_map}, we insert the operator $e^{-s \hat{k}}$ [upper part of Fig.~\ref{fig:TN_diagram}(d)] in the contracted ancilla legs of the previous transfer matrix [Fig.~\ref{fig:TN_diagram}(e)], with $\hat{k}=\ketbra{1}$ the projector onto the excited state of the ancilla, which ensures the appropriate $k-$dependent weight. Conditioning on a particular measurement record  amounts instead to inserting specific projectors on the ancilla indices [lower part of Fig.~\ref{fig:TN_diagram}(d)] at each collision step. The probability of a given trajectory is obtained by tracing over the system indices after applying the corresponding sequence of projectors [Fig.~\ref{fig:TN_diagram}(f)], and is equivalent to an expectation value in the MPS~\cite{kiukas_2015_cMPS}.

The construction can be extended for a many-body system with local interactions, such as the one described by the Hamiltonian defined in Eq.~\eqref{eq:Rydberg_Hamiltonian}. In one spatial dimension, the state of the system is represented as a matrix product operator (MPO), and the unitary part of the quantum channel can be written as a product of MPOs using a Trotter expansion (see SM for details~\cite{SM}). Local measurement of ancillas and  bias operators given by sums of single site observables preserve this structure and can be added to the ancilla legs as in the single system case described above [Fig.~\ref{fig:TN_diagram}(g)]. In our case, the bias operator takes the form $e^{-s \hat{A}}$, where the activity operator for a single collision step is $\hat{A} = \sum_{i=1}^{L} \hat{k}_{i}$. Allowing for an MPO structure of the bias, the framework can accommodate not only activity bias but also space, time, and space–time correlations~\cite{Cech_2025}. 

The problem of finding the dominant eigenvalue of the tilted $\mathcal{E}_s$ can be solved within the TN framework via a power method: applying the MPO for $\mathcal{E}_s$ to the system state produces another MPO of larger bond dimension, which is then truncated to a bounded one. Repeating this procedure eventually converges and provides an estimate of the dominant eigenvalue (and its associated right eigenvector) and thus the SCGF $\theta(s)$, up to the error induced by the Trotterization and the truncation. Because the application and truncation of the MPO is efficient in terms of the system size, large systems become accessible. In turn, the numerical errors need to be systematically controlled to ensure convergence (see SM for details~\cite{SM}).

\textbf{Dynamical phase coexistence and glassiness. ---}
In the following we use the TN representation of trajectory ensembles to explore the dynamical phases of the model defined by Eqs.~(\ref{eq:collision_hamiltonian}) and (\ref{eq:Rydberg_Hamiltonian}), fixing $\Omega = 1.0$, $\gamma/\Omega = 3.0$, and $\Omega  \Delta t = 1.25$, while varying the interaction strength $V/\Omega $, the counting field $s$, and the system size $L$. Figure~\ref{fig:numerics} shows, for two representative~\cite{Cech_2025} values of $V/\Omega $, the SCGF [panel (a)], the activity [panel (b)] and the rate function [panel (c)]. For $V/\Omega =2$, the SCGF and activity remain smooth functions of $s$, while the rate function $-\phi(a)$ exhibits a unique maximum, with almost no dependence on system size. In contrast, for $V/\Omega =5.875$, the SCGF develops a non-analyticity close to $s=0$, and the activity shows a rapidly sharpening crossover as $L$ increases, consistent with a first-order dynamical phase transition~\cite{lecomte_thermodynamic_2007,garrahan_first-order_2009}. At the same time, the rate function broadens, and develops a flat region, consistent with the Maxwell construction on a bimodal distribution, with peaks at $a \approx 0.5$ and $a \approx 0$, reflecting the coexistence of active and inactive trajectories~\cite{lecomte_thermodynamic_2007,garrahan_dynamical_2007}. 

To locate the transition more precisely, we analyze the finite-size scaling of the crossing of $\theta(s)$ in Fig.~\ref{fig:numerics} (d,e). For $V/\Omega=5.875$ [Fig.~\ref{fig:numerics}(e)], in the inactive phase ($s \gg 0$) $\theta(s)$ is close to zero and nearly constant, whereas near $s\approx 0$ first-order perturbation theory predicts a linear behavior with slope $-a_0$, where $a_0$ is the mean activity at $s=0$. The crossing point of these two branches shifts towards $s=0$ as $L$ increases. Extrapolating the finite-size estimates with $1/L$ (inset) yields $s^* = (2 \pm 3) \times 10^{-5}$, consistent with a transition at $s=0$ in the thermodynamic limit. This indicates that, for these parameters, the original dynamics displays phase coexistence between active and inactive trajectories. Such dynamical heterogeneity, when persisting in the thermodynamic limit, is characteristic for glass formers~\cite{garrahan_dynamical_2007,hedges_dynamic_2009,chandler_dynamics_2010}.

Repeating the analysis for other $V/\Omega $ values, we map out the phase diagram in $(V/\Omega ,s)$ space, shown in Fig.~\ref{fig:phase_diagram}(c). The extrapolated $s^*$ values (grey crosses) delineate the first-order phase boundary between active (dark) and inactive (light) regimes. Transitions at $s^*=0$ (red crosses) correspond to phase coexistence in the original dynamics, which we identify at $V/\Omega =4.3,$ $5.875$ and $6.8$. Transitions at other values $s^*\neq 0$ reflect instead singularities in the statistics of rare trajectories~\cite{lecomte_thermodynamic_2007,garrahan_dynamical_2007,garrahan_first-order_2009}.

\begin{figure}[t]
    \centering
    \includegraphics[width=\columnwidth]{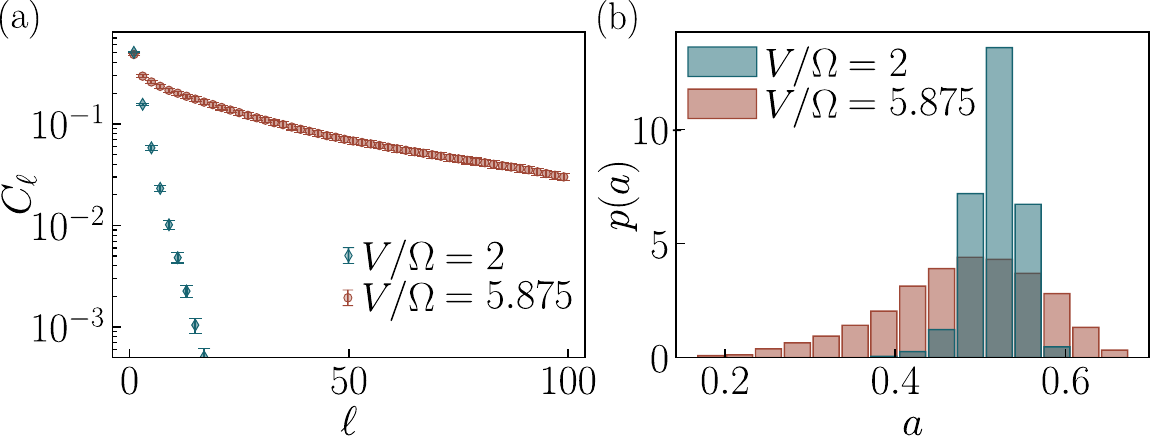}
    \vspace{-20pt}
    \caption{\textbf{String correlator in the conditioned ensemble.} (a) String correlator [Eq.~\eqref{eq:string_correl}] as a function of the string length $\ell$ for $L=20$, averaged over the two central sites of the chain. Results are shown for $V/\Omega=2$ (blue diamonds) and $V/\Omega=5.875$ (red circles). (b) Probability distribution of the activity $p(a)$ obtained from 1000 sampled trajectories, showing enhanced weight at low activity in the coexistence regime.
   \label{fig:conditioned}
   }    
\end{figure}

\textbf{Underlying conditioned quantum dynamics. --}
So far we have investigated the statistics of the trajectories constructed from ancilla measurements. A natural question is whether the dynamical regimes identified in trajectory space have a direct correspondence in the quantum state of the system and not only in the measurement record. This question can be addressed in the conditioned ensemble. The TN structure of the probability allows efficient sampling of outcomes~\cite{ferris_2012_perfect,iblisdir_2014_annealing} and evaluation of expectation values in the conditioned state of the system along the given trajectory. To probe the persistence of \emph{inactive} regions in the measurement record, corresponding to periods where system sites remain close to the excited state~\cite{cech_2025_revealing}, we introduce the following string correlator
\begin{align} \label{eq:string_correl}
C_{\ell}^{i}\equiv  \sum_{\eta(T)} \pi(\eta(T))
\prod_{m=1}^{\ell}\expval{\hat{n}_i(m)}_{\eta},       
\end{align}
where $\expval{\hat{n}_i(m)}_{\eta}$ denotes the expected occupation of site $i$ at collision step $m$, conditioned to outcome record $\eta(m)$~\footnote{Strictly speaking, the expectation value $\expval{\hat{n}_i(m)}_{\eta}$ is conditioned only on the measurement outcomes up to collision step $m$, i.e., on the partial trajectory $\eta(m)$, rather than on the full trajectory $\eta(T)$.}. Because it is a non-linear functional of the measurement record, $C_{\ell}$ is not equivalent to an observable in the average ensemble. Since the product becomes large only when a site remains excited for $\ell$ consecutive collision steps, the correlator detects trajectories with long-lived inactive regions, even if occurring with low probability. As illustrated in Fig.~\ref{fig:conditioned}(a) for $L=20$, $C_{\ell}$ decays fast for $V/\Omega=2$, consistent with inactive regions that live shortly, but much more slowly for $V/\Omega=5.875$, indicating persistent inactive regions, in correspondence with the enhanced probability of low-activity trajectories [Fig.~\ref{fig:conditioned}(b)]. This shows that the dynamical coexistence identified in trajectory space is directly encoded in the conditioned quantum states of the system. The phase transition is therefore not merely a feature of the measurement statistics, but is imprinted in the quantum dynamics itself.

\textbf{Conclusions. ---}
We have introduced and demonstrated a TN framework for the LD analysis of open quantum systems with discrete-time dynamics. Our approach enables the scalable computation of dynamical observables and the detection of phase transitions in trajectory space. Furthermore it provides access to the conditioned ensemble of (monitored) system evolutions. Applying this method to a quantum many-body system inspired by Rydberg gases, we have established the presence of a first-order dynamical phase transition between active and inactive phases. The associated phase diagram features several points of dynamical phase coexistence, which is a manifestation of glassy dynamics. Our approach combines two complementary perspectives: we identify the transition in the statistics of measurement records (trajectory space), using the LD formalism,  and probe the corresponding  system dynamics through the conditioned observables, thus providing a microscopic interpretation of the active-inactive phase coexistence. The method can be readily generalized to other models and biasing observables, widening the possibility of studying rare-event statistics and nonequilibrium behavior in quantum devices. And, even though less immediate, an interesting direction is the extension to models with long-range interactions, as naturally realized in Rydberg platforms.

The code and data that support the findings of this work are available on Zenodo~\cite{Cea_Fernandez_Data_and_code_2026}.


\acknowledgments
\textit{Acknowledgements.---} We thank R. Vasseur for fruitful discussions. We acknowledge funding from the Deutsche Forschungsgemeinschaft (DFG, German Research Foundation) under Germany's Excellence Strategy -- EXC-2111 -- 390814868, through the Research Unit FOR 5413/1, Grant No. 465199066, and through the Research Unit FOR 5522/1, Grant No. 499180199. This work is supported by ERC grant OPEN-2QS (Grant No. 101164443, https://doi.org/10.3030/101164443).

\bibliography{biblio}

@article{Cech_2025,
  title = {Space-Time Correlations in Monitored Kinetically Constrained Discrete-Time Quantum Dynamics},
  author = {Cech, Marcel and Cea, Mar\'{\i}a and Ba\~nuls, Mari Carmen and Lesanovsky, Igor and Carollo, Federico},
  journal = {Phys. Rev. Lett.},
  volume = {134},
  issue = {23},
  pages = {230403},
  numpages = {9},
  year = {2025},
  month = {Jun},
  publisher = {American Physical Society},
  doi = {10.1103/2lxs-wccj},
  url = {https://link.aps.org/doi/10.1103/2lxs-wccj}
}

@article{cech_2025_revealing,
      title={Revealing emergent many-body phenomena by analyzing large-scale space-time records of monitored quantum systems}, 
      author={Cech, Marcel and De Fazio, Cecilia and Cea, María and Bañuls, Mari Carmen and Lesanovsky, Igor and Carollo, Federico},
      year={2025},
      journal={arXiv:2507.00944},
      archivePrefix={arXiv},
      primaryClass={quant-ph},
      url={https://arxiv.org/abs/2507.00944}, 
}

@article{ciccarello_quantum_2022,
	title = {Quantum collision models: {Open} system dynamics from repeated interactions},
	volume = {954},
	issn = {0370-1573},
	shorttitle = {Quantum collision models},
	url = {https://www.sciencedirect.com/science/article/pii/S0370157322000035},
	doi = {10.1016/j.physrep.2022.01.001},
	urldate = {2025-09-14},
	journal = {Phys. Rep.},
	author = {Ciccarello, Francesco and Lorenzo, Salvatore and Giovannetti, Vittorio and Palma, G. Massimo},
	month = apr,
	year = {2022},
	pages = {1--70},
}

@misc{nielsen_quantum_2010,
	title = {Quantum {Computation} and {Quantum} {Information}: 10th {Anniversary} {Edition}},
	shorttitle = {Quantum {Computation} and {Quantum} {Information}},
	url = {https://www.cambridge.org/highereducation/books/quantum-computation-and-quantum-information/01E10196D0A682A6AEFFEA52D53BE9AE},
	urldate = {2025-09-14},
	journal = {Cambridge Aspire website},
	author = {Nielsen, Michael A. and Chuang, Isaac L.},
	month = dec,
	year = {2010},
	doi = {10.1017/CBO9780511976667},
	note = {ISBN: 9780511976667
Publisher: Cambridge University Press},
}

@article{saffman_quantum_2010,
	title = {Quantum information with {Rydberg} atoms},
	volume = {82},
	doi = {10.1103/RevModPhys.82.2313},
	number = {3},
	journal = {Rev. Mod. Phys.},
	author = {Saffman, M.},
	year = {2010},
	pages = {2313--2363},
}

@article{bernien_probing_2017,
	title = {Probing many-body dynamics on a 51-atom quantum simulator},
	volume = {551},
	copyright = {2017 Macmillan Publishers Limited, part of Springer Nature. All rights reserved.},
	issn = {1476-4687},
	url = {https://www.nature.com/articles/nature24622},
	doi = {10.1038/nature24622},
	number = {7682},
	urldate = {2025-09-14},
	journal = {Nature},
	author = {Bernien, Hannes and Schwartz, Sylvain and Keesling, Alexander and Levine, Harry and Omran, Ahmed and Pichler, Hannes and Choi, Soonwon and Zibrov, Alexander S. and Endres, Manuel and Greiner, Markus and Vuletić, Vladan and Lukin, Mikhail D.},
	month = nov,
	year = {2017},
	pages = {579--584},
}

@article{lesanovsky_interacting_2012,
  title = {Interacting Fibonacci anyons in a Rydberg gas},
  author = {Lesanovsky, Igor and Katsura, Hosho},
  journal = {Phys. Rev. A},
  volume = {86},
  issue = {4},
  pages = {041601},
  numpages = {5},
  year = {2012},
  month = {Oct},
  publisher = {American Physical Society},
  doi = {10.1103/PhysRevA.86.041601},
  url = {https://link.aps.org/doi/10.1103/PhysRevA.86.041601}
}

@article{turner_quantum_2018,
	title = {Quantum scarred eigenstates in a {Rydberg} atom chain: {Entanglement}, breakdown of thermalization, and stability to perturbations},
	volume = {98},
	shorttitle = {Quantum scarred eigenstates in a {Rydberg} atom chain},
	doi = {10.1103/PhysRevB.98.155134},
	number = {15},
	journal = {Phys. Rev. B},
    pages = {155134},
	author = {Turner, C. J.},
	year = {2018},
}

@article{moudgalya_quantum_2022,
	title = {Quantum many-body scars and {Hilbert} space fragmentation: a review of exact results},
	volume = {85},
	issn = {0034-4885},
	shorttitle = {Quantum many-body scars and {Hilbert} space fragmentation},
	url = {https://dx.doi.org/10.1088/1361-6633/ac73a0},
	doi = {10.1088/1361-6633/ac73a0},
	number = {8},
	urldate = {2025-09-14},
	journal = {Reports on Progress in Physics},
	author = {Moudgalya, Sanjay and Bernevig, B Andrei and Regnault, Nicolas},
	month = jul,
	year = {2022},
	pages = {086501},
}

@article{serbyn_quantum_2021,
	title = {Quantum many-body scars and weak breaking of ergodicity},
	volume = {17},
	copyright = {2021 Springer Nature Limited},
	issn = {1745-2481},
	url = {https://www.nature.com/articles/s41567-021-01230-2},
	doi = {10.1038/s41567-021-01230-2},
	number = {6},
	urldate = {2025-09-14},
	journal = {Nat. Phys.},
	author = {Serbyn, Maksym and Abanin, Dmitry A. and Papić, Zlatko},
	month = jun,
	year = {2021},
	pages = {675--685},
}

@article{touchette_large_2009,
	title = {The large deviation approach to statistical mechanics},
	volume = {478},
	issn = {0370-1573},
	url = {https://www.sciencedirect.com/science/article/pii/S0370157309001410},
	doi = {10.1016/j.physrep.2009.05.002},
	number = {1},
	urldate = {2025-09-16},
	journal = {Phys. Rep.},
	author = {Touchette, Hugo},
	month = jul,
	year = {2009},
	pages = {1--69},
}

@article{garrahan_thermodynamics_2010,
  title = {Thermodynamics of Quantum Jump Trajectories},
  author = {Garrahan, Juan P. and Lesanovsky, Igor},
  journal = {Phys. Rev. Lett.},
  volume = {104},
  issue = {16},
  pages = {160601},
  numpages = {4},
  year = {2010},
  month = {Apr},
  publisher = {American Physical Society},
  doi = {10.1103/PhysRevLett.104.160601},
  url = {https://link.aps.org/doi/10.1103/PhysRevLett.104.160601}
}

@article{hedges_dynamic_2009,
	title = {Dynamic {Order}-{Disorder} in {Atomistic} {Models} of {Structural} {Glass} {Formers}},
	volume = {323},
	url = {https://www.science.org/doi/10.1126/science.1166665},
	doi = {10.1126/science.1166665},
	number = {5919},
	urldate = {2025-09-16},
	journal = {Science},
	author = {Hedges, Lester O. and Jack, Robert L. and Garrahan, Juan P. and Chandler, David},
	month = mar,
	year = {2009},
	pages = {1309--1313},
}

@article{schollwock_density-matrix_2011,
	title = {The density-matrix renormalization group in the age of matrix product states},
	volume = {326},
	issn = {0003-4916},
	url = {https://www.sciencedirect.com/science/article/pii/S0003491610001752},
	doi = {10.1016/j.aop.2010.09.012},
	number = {1},
	urldate = {2024-01-14},
	journal = {Ann. Phys.},
	author = {Schollwöck, Ulrich},
	month = jan,
	year = {2011},
	pages = {96--192},
}

@article{orus_practical_2014,
	title = {A {Practical} {Introduction} to {Tensor} {Networks}: {Matrix} {Product} {States} and {Projected} {Entangled} {Pair} {States}},
	volume = {349},
	issn = {00034916},
	shorttitle = {A {Practical} {Introduction} to {Tensor} {Networks}},
	url = {http://arxiv.org/abs/1306.2164},
	doi = {10.1016/j.aop.2014.06.013},
	urldate = {2024-01-14},
	journal = {Ann. Phys.},
	author = {Orus, Roman},
	month = oct,
	year = {2014},
	pages = {117--158},
}

@article{banuls_tensor_2023,
	title = {Tensor {Network} {Algorithms}: {A} {Route} {Map}},
	volume = {14},
	issn = {1947-5454, 1947-5462},
	shorttitle = {Tensor {Network} {Algorithms}},
	url = {https://www.annualreviews.org/content/journals/10.1146/annurev-conmatphys-040721-022705},
	doi = {10.1146/annurev-conmatphys-040721-022705},
	number = {Volume 14, 2023},
	urldate = {2025-10-13},
	journal = {Annu. Rev. Condens. Matter Phys.},
	author = {Bañuls, Mari Carmen},
	month = mar,
	year = {2023},
	pages = {173--191},
}

@article{ritort_glassy_2003,
	title = {Glassy dynamics of kinetically constrained models},
	volume = {52},
	issn = {0001-8732},
	url = {https://doi.org/10.1080/0001873031000093582},
	doi = {10.1080/0001873031000093582},
	number = {4},
	urldate = {2025-10-13},
	journal = {Adv. Phys.},
	author = {Ritort, F. and Sollich, P.},
	month = jun,
	year = {2003},
	pages = {219--342},
}

@article{garrahan_dynamical_2007,
	title = {Dynamical {First}-{Order} {Phase} {Transition} in {Kinetically} {Constrained} {Models} of {Glasses}},
	volume = {98},
	doi = {10.1103/PhysRevLett.98.195702},
	number = {19},
	journal = {Phys. Rev. Lett.},
	author = {Garrahan, J. P. and Jack, R. L. and Lecomte, V. and Pitard, E. and van Duijvendijk, K. and van Wijland, F.},
	year = {2007},
    pages={195702},
}

@article{garrahan_first-order_2009,
	title = {First-order dynamical phase transition in models of glasses: an approach based on ensembles of histories},
	volume = {42},
	issn = {1751-8121},
	shorttitle = {First-order dynamical phase transition in models of glasses},
	url = {https://doi.org/10.1088/1751-8113/42/7/075007},
	doi = {10.1088/1751-8113/42/7/075007},
	number = {7},
	urldate = {2025-10-13},
	journal = {J. Phys. A Math. Theor.},
	author = {Garrahan, Juan P and Jack, Robert L and Lecomte, Vivien and Pitard, Estelle and van Duijvendijk, Kristina and van Wijland, Frédéric},
	month = jan,
	year = {2009},
	pages = {075007},
}

@article{lecomte_thermodynamic_2007,
	title = {Thermodynamic {Formalism} for {Systems} with {Markov} {Dynamics}},
	volume = {127},
	issn = {1572-9613},
	url = {https://doi.org/10.1007/s10955-006-9254-0},
	doi = {10.1007/s10955-006-9254-0},
	number = {1},
	urldate = {2025-10-13},
	journal = {J. Stat. Phys.},
	author = {Lecomte, V. and Appert-Rolland, C. and van Wijland, F.},
	month = apr,
	year = {2007},
	pages = {51--106},
}

@article{Sala_2020_hsf,
  title = {Ergodicity Breaking Arising from Hilbert Space Fragmentation in Dipole-Conserving Hamiltonians},
  author = {Sala, Pablo and Rakovszky, Tibor and Verresen, Ruben and Knap, Michael and Pollmann, Frank},
  journal = {Phys. Rev. X},
  volume = {10},
  issue = {1},
  pages = {011047},
  numpages = {19},
  year = {2020},
  month = {Feb},
  publisher = {American Physical Society},
  doi = {10.1103/PhysRevX.10.011047},
  url = {https://link.aps.org/doi/10.1103/PhysRevX.10.011047}
}

@article{barratt_2022_learn,
  title = {Transitions in the Learnability of Global Charges from Local Measurements},
  author = {Barratt, Fergus and Agrawal, Utkarsh and Potter, Andrew C. and Gopalakrishnan, Sarang and Vasseur, Romain},
  journal = {Phys. Rev. Lett.},
  volume = {129},
  issue = {20},
  pages = {200602},
  numpages = {7},
  year = {2022},
  month = {Nov},
  publisher = {American Physical Society},
  doi = {10.1103/PhysRevLett.129.200602},
  url = {https://link.aps.org/doi/10.1103/PhysRevLett.129.200602}
}

@article{ippolity_2024_learn,
  title = {Learnability Transitions in Monitored Quantum Dynamics via Eavesdropper's Classical Shadows},
  author = {Ippoliti, Matteo and Khemani, Vedika},
  journal = {PRX Quantum},
  volume = {5},
  issue = {2},
  pages = {020304},
  numpages = {24},
  year = {2024},
  month = {Apr},
  publisher = {American Physical Society},
  doi = {10.1103/PRXQuantum.5.020304},
  url = {https://link.aps.org/doi/10.1103/PRXQuantum.5.020304}
}

@article{gopalakrishnan_2026_hydro,
  title = {Monitored fluctuating hydrodynamics},
  author = {Gopalakrishnan, Sarang and McCulloch, Ewan and Vasseur, Romain},
  journal = {Phys. Rev. X},
  volume = {16},
  pages = {011024},
  year = {2026},
  month = {Jan},
  publisher = {American Physical Society},
  doi = {10.1103/295c-lj1w},
  url = {https://link.aps.org/doi/10.1103/295c-lj1w}
}

@article{nahum_2025_bayesian,
  title = {Bayesian critical points in classical lattice models},
  author = {Nahum, Adam and Jacobsen, Jesper Lykke},
  journal = {Phys. Rev. B},
  volume = {112},
  issue = {23},
  pages = {235113},
  numpages = {57},
  year = {2025},
  month = {Dec},
  publisher = {American Physical Society},
  doi = {10.1103/7dpt-d4s5},
  url = {https://link.aps.org/doi/10.1103/7dpt-d4s5}
}

@article{causer2021optimal,
  title = {Optimal sampling of dynamical large deviations via matrix product states},
  author = {Causer, Luke and Ba\~nuls, Mari Carmen and Garrahan, Juan P.},
  journal = {Phys. Rev. E},
  volume = {103},
  issue = {6},
  pages = {062144},
  numpages = {12},
  year = {2021},
  month = {Jun},
  publisher = {American Physical Society},
  doi = {10.1103/PhysRevE.103.062144},
  url = {https://link.aps.org/doi/10.1103/PhysRevE.103.062144}
}

@article{causer2023optimal,
  title = {Optimal Sampling of Dynamical Large Deviations in Two Dimensions via Tensor Networks},
  author = {Causer, Luke and Ba\~nuls, Mari Carmen and Garrahan, Juan P.},
  journal = {Phys. Rev. Lett.},
  volume = {130},
  issue = {14},
  pages = {147401},
  numpages = {6},
  year = {2023},
  month = {Apr},
  publisher = {American Physical Society},
  doi = {10.1103/PhysRevLett.130.147401},
  url = {https://link.aps.org/doi/10.1103/PhysRevLett.130.147401}
}

@article{banuls2019using,
  title={Using matrix product states to study the dynamical large deviations of kinetically constrained models},
  author={Banuls, Mari Carmen and Garrahan, Juan P},
  journal={Phys. Rev. Lett.},
  volume={123},
  number={20},
  pages={200601},
  year={2019},
  publisher={APS},
  doi       = {10.1103/PhysRevLett.123.200601},
  url       = {https://link.aps.org/doi/10.1103/PhysRevLett.123.200601},
}

@article{olmos2014out,
  title = {Out-of-equilibrium evolution of kinetically constrained many-body quantum systems under purely dissipative dynamics},
  author = {Olmos, Beatriz and Lesanovsky, Igor and Garrahan, Juan P.},
  journal = {Phys. Rev. E},
  volume = {90},
  issue = {4},
  pages = {042147},
  numpages = {6},
  year = {2014},
  month = {Oct},
  publisher = {American Physical Society},
  doi = {10.1103/PhysRevE.90.042147},
  url = {https://link.aps.org/doi/10.1103/PhysRevE.90.042147}
}

@article{ostmann2019localization,
  title = {Localization in spin chains with facilitation constraints and disordered interactions},
  author = {Ostmann, Maike and Marcuzzi, Matteo and Garrahan, Juan P. and Lesanovsky, Igor},
  journal = {Phys. Rev. A},
  volume = {99},
  issue = {6},
  pages = {060101},
  numpages = {7},
  year = {2019},
  month = {Jun},
  publisher = {American Physical Society},
  doi = {10.1103/PhysRevA.99.060101},
  url = {https://link.aps.org/doi/10.1103/PhysRevA.99.060101}
}

@article{scherg2021observing,
  title={Observing non-ergodicity due to kinetic constraints in tilted Fermi-Hubbard chains},
  author={Scherg, Sebastian and Kohlert, Thomas and Sala, Pablo and Pollmann, Frank and Hebbe Madhusudhana, Bharath and Bloch, Immanuel and Aidelsburger, Monika},
  journal={Nat. Commun.},
  volume={12},
  number={1},
  pages={4490},
  year={2021},
  publisher={Nature Publishing Group UK London},
  doi = {10.1038/s41467-021-24726-0},
  url = {https://www.nature.com/articles/s41467-021-24726-0}
}

@article{kiukas_2015_cMPS,
  title = {Equivalence of matrix product ensembles of trajectories in open quantum systems},
  author = {Kiukas, Jukka and Guţă, Mădălin and Lesanovsky, Igor and Garrahan, Juan P.},
  journal = {Phys. Rev. E},
  volume = {92},
  issue = {1},
  pages = {012132},
  numpages = {11},
  year = {2015},
  month = {Jul},
  publisher = {American Physical Society},
  doi = {10.1103/PhysRevE.92.012132},
  url = {https://link.aps.org/doi/10.1103/PhysRevE.92.012132}
}

@article{ferris_2012_perfect,
  title = {Perfect sampling with unitary tensor networks},
  author = {Ferris, Andrew J. and Vidal, Guifre},
  journal = {Phys. Rev. B},
  volume = {85},
  issue = {16},
  pages = {165146},
  numpages = {10},
  year = {2012},
  month = {Apr},
  publisher = {American Physical Society},
  doi = {10.1103/PhysRevB.85.165146},
  url = {https://link.aps.org/doi/10.1103/PhysRevB.85.165146}
}

@article{iblisdir_2014_annealing,
	author = {Iblisdir, S},
	journal = {New J. Phys.},
	month = {oct},
	number = {10},
	pages = {103022},
	title = {Simulated annealing for tensor network states},
	volume = {16},
	year = {2014},
    doi = {10.1088/1367-2630/16/10/103022}
}

@article{Gorissen2009dmrg,
  title = {Density-matrix renormalization-group study of current and activity fluctuations near nonequilibrium phase transitions},
  author = {Gorissen, Mieke and Hooyberghs, Jef and Vanderzande, Carlo},
  journal = {Phys. Rev. E},
  volume = {79},
  issue = {2},
  pages = {020101},
  numpages = {4},
  year = {2009},
  month = {Feb},
  publisher = {American Physical Society},
  doi = {10.1103/PhysRevE.79.020101},
  url = {https://link.aps.org/doi/10.1103/PhysRevE.79.020101}
}

@article{browaeys2020manybody,
  title = {Many-body physics with individually controlled Rydberg atoms},
  author = {Browaeys, A. and Lahaye, T.},
  journal = {Nat. Phys.},
  volume = {16},
  pages = {132--142},
  year = {2020},
  doi={https://doi.org/10.1038/s41567-019-0733-z}
}

@article{labuhn2016tunable,
  title = {Tunable two-dimensional arrays of single Rydberg atoms for realizing quantum Ising models},
  author = {Labuhn, H. and Barredo, D. and Ravets, S. and de L{\'e}s{\'e}leuc, S. and Macr{\`\i}, T. and Lahaye, T. and Browaeys, A.},
  journal = {Nature},
  volume = {534},
  pages = {667--670},
  year = {2016},
  doi = {https://doi.org/10.1038/nature18274}
}

@article{ebadi2021quantum,
  title = {Quantum phases of matter on a 256-atom programmable quantum simulator},
  author = {Ebadi, Sepehr and Wang, Tout T. and Levine, Harry and Keesling, Alexander and Semeghini, Giulia and Omran, Ahmed and Bluvstein, Dolev and Samajdar, Rhine and Pichler, Hannes and Ho, Wen Wei and Choi, Soonwon and Sachdev, Subir and Greiner, Markus and Vuletić, Vladan and Lukin, Mikhail D.},
  journal = {Nature},
  volume = {595},
  pages = {227--232},
  year = {2021},
  doi = {https://doi.org/10.1038/s41586-021-03582-4}
}

@article{Li2018,
  title = {Quantum Zeno effect and the many-body entanglement transition},
  author = {Li, Yaodong and Chen, Xiao and Fisher, Matthew P. A.},
  journal = {Phys. Rev. B},
  volume = {98},
  issue = {20},
  pages = {205136},
  numpages = {9},
  year = {2018},
  month = {Nov},
  publisher = {American Physical Society},
  doi = {10.1103/PhysRevB.98.205136},
  url = {https://link.aps.org/doi/10.1103/PhysRevB.98.205136}
}

@article{LiChenFisher2019,
  title = {Measurement-driven entanglement transition in hybrid quantum circuits},
  author = {Li, Yaodong and Chen, Xiao and Fisher, Matthew P. A.},
  journal = {Phys. Rev. B},
  volume = {100},
  issue = {13},
  pages = {134306},
  numpages = {26},
  year = {2019},
  month = {Oct},
  publisher = {American Physical Society},
  doi = {10.1103/PhysRevB.100.134306},
  url = {https://link.aps.org/doi/10.1103/PhysRevB.100.134306}
}

@article{Skinner2019,
  title = {Measurement-Induced Phase Transitions in the Dynamics of Entanglement},
  author = {Skinner, Brian and Ruhman, Jonathan and Nahum, Adam},
  journal = {Phys. Rev. X},
  volume = {9},
  issue = {3},
  pages = {031009},
  numpages = {21},
  year = {2019},
  month = {Jul},
  publisher = {American Physical Society},
  doi = {10.1103/PhysRevX.9.031009},
  url = {https://link.aps.org/doi/10.1103/PhysRevX.9.031009}
}

@article{DAlessio2016,
  title = {From quantum chaos and eigenstate thermalization to statistical mechanics and thermodynamics},
  author = {D'Alessio, Luca and Kafri, Yariv and Polkovnikov, Anatoli and Rigol, Marcos},
  journal = {Adv. Phys.},
  volume = {65},
  number = {3},
  pages = {239--362},
  year = {2016},
  doi = {10.1080/00018732.2016.1198134}
}

@article{FredricksonAndersen1984,
  title = {Kinetic Ising Model of the Glass Transition},
  author = {Fredrickson, Glenn H. and Andersen, Hans C.},
  journal = {Phys. Rev. Lett.},
  volume = {53},
  issue = {13},
  pages = {1244--1247},
  numpages = {0},
  year = {1984},
  month = {Sep},
  publisher = {American Physical Society},
  doi = {10.1103/PhysRevLett.53.1244},
  url = {https://link.aps.org/doi/10.1103/PhysRevLett.53.1244}
}

@article{Daley2014,
author = {Andrew J. Daley},
title = {Quantum trajectories and open many-body quantum systems},
journal = {Adv. Phys.},
volume = {63},
number = {2},
pages = {77--149},
year = {2014},
publisher = {Taylor \& Francis},
doi = {10.1080/00018732.2014.933502},
}

@book{goldenfeld2018lectures,
  title={Lectures on phase transitions and the renormalization group},
  author={Goldenfeld, Nigel},
  year={2018},
  publisher={CRC Press},
  doi={https://doi.org/10.1201/9780429493492}
}

@misc{SM,
  title = {See Supplemental Material},
  note = {See Supplemental Material for details on the Trotter decomposition, convergence with bond dimension, exact MPO computation of the activity, and estimation of the critical counting field $s^*$.}
}

@book{Sakurai_Napolitano_2020, place={Cambridge}, edition={3}, title={Modern Quantum Mechanics}, publisher={Cambridge University Press}, author={Sakurai, J. J. and Napolitano, Jim}, year={2020}, doi={
https://doi.org/10.1017/9781108587280}}

@article{chandler_dynamics_2010,
	title = {Dynamics on the {Way} to {Forming} {Glass}: {Bubbles} in {Space}-{Time}},
	volume = {61},
	issn = {0066-426X, 1545-1593},
	shorttitle = {Dynamics on the {Way} to {Forming} {Glass}},
	url = {https://www.annualreviews.org/content/journals/10.1146/annurev.physchem.040808.090405},
	doi = {10.1146/annurev.physchem.040808.090405},
	number = {Volume 61, 2010},
	urldate = {2026-03-19},
	journal = {Annual Review of Physical Chemistry},
	publisher = {Annual Reviews},
	author = {Chandler, David and Garrahan, Juan P.},
	month = may,
	year = {2010},
	pages = {191--217},
}

@misc{Cea_Fernandez_Data_and_code_2026,
author = {Cea Fernández, María and Cech, Marcel and Carollo, Federico and Lesanovsky, Igor and Bañuls, Mari Carmen},
doi = {10.5281/zenodo.19207105},
month = mar,
title = {{Data and code for Large deviations and conditioned monitored quantum systems: a tensor network approach}},
url = {https://doi.org/10.5281/zenodo.19207104},
version = {v0.1.0},
year = {2026}
}

\onecolumngrid
\newpage

\setcounter{equation}{0}
\setcounter{page}{1}

\setcounter{figure}{0}
\setcounter{table}{0}
\makeatletter
\renewcommand{\theequation}{S\arabic{equation}}
\renewcommand{\thefigure}{S\arabic{figure}}
\renewcommand{\thetable}{S\arabic{table}}
\setcounter{secnumdepth}{1}

\begin{center}
{\Large SUPPLEMENTAL MATERIAL}
\end{center}
\begin{center}
\vspace{0.8cm}
{\Large Large deviations and conditioned monitored quantum systems: a tensor network approach}
\end{center}
\begin{center}
María Cea$^{1,2}$, Marcel Cech$^{3}$, Federico Carollo$^{4,5}$, Igor Lesanovsky$^{3,6}$, Mari Carmen Bañuls$^{1,2}$
\end{center}
\begin{center}
$^1${\em Max-Plank-Institut f\"ur Quantenoptik, Hans-Kopfermann-Str. 1, D-85748 Garching, Germany}\\
$^2${\em Munich Center for Quantum Science and Technology (MCQST), Schellingstr. 4, D-80799 M\"unchen, Germany}\\
$^3${\em Institut f\"ur Theoretische Physik and Center for Integrated Quantum Science and Technology,\\ Universit\"at T\"ubingen, Auf der Morgenstelle 14, 72076 T\"ubingen, Germany}\\
$^4${\em Dipartimento di Fisica, Sapienza Università di Roma, Piazzale Aldo Moro 2, 00185 Rome, Italy}\\
$^5${\em Centre for Fluid and Complex Systems, Coventry University, Coventry, CV1 2TT, United Kingdom}\\
$^6${\em School of Physics and Astronomy and Centre for the Mathematics and Theoretical Physics of Quantum Non-Equilibrium Systems, The University of Nottingham, Nottingham, NG7 2RD, United Kingdom}
\end{center}

\section{Error analysis}
The accuracy of the TN method employed to compute the eigenvalues in \eqref{eq:biased_map} is limited by two controlled approximations: the Trotter decomposition of the unitary system–ancilla evolution used to construct the MPO channel, and the truncation of the bond dimension in the approximate eigenvector.

\begin{figure}[h]
    \centering
    \includegraphics[width=0.8\textwidth]{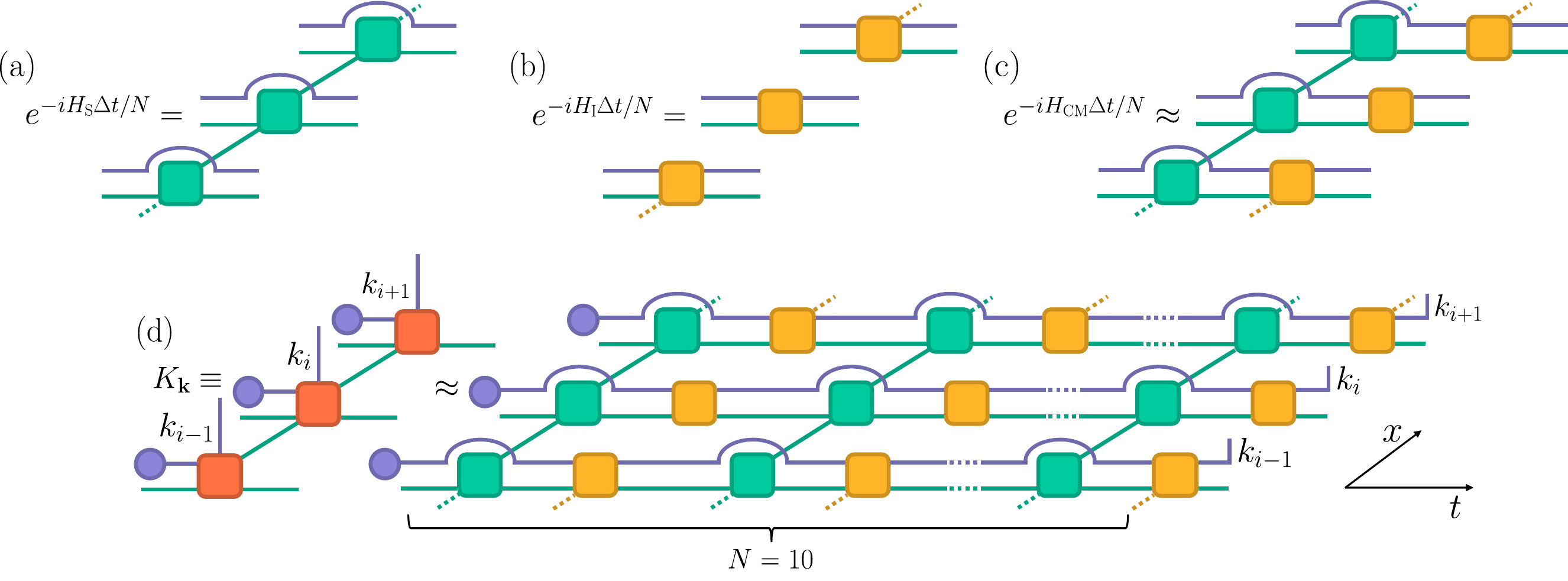}
    \caption{\textbf{Trotterization.} (a) MPO representation of $e^{-i H_{\mathrm{S}} \Delta t / N}$ obtained from the even–odd decomposition (bond dimension 2). (b) MPO representation of $e^{-i H_{\mathrm{I}} \Delta t / N}$ with product structure (bond dimension 1). (c) MPO for a single Trotter step obtained by contracting (a) and (b). (d) Kraus operator constructed as the contraction of $N$ Trotter steps.}
    \label{fig:trotterization}
\end{figure}

\subsection{Trotterization}
The Kraus operators defining the dynamics of each collision step are constructed from a unitary operator $U_{\mathrm{CM}}=e^{-i H_{\mathrm{CM}} \Delta t} $. The collision model Hamiltonian
\begin{align}
    H_\mathrm{CM} =  H_\mathrm{S} \otimes \Id + H_\mathrm{I} \, ,
    \label{Seq:collision_hamiltonian}
\end{align}
with
\begin{align}
    H_\mathrm{S} = \Omega \sum_{i=1}^L \sigma_i^x + V \sum_{i=1}^{L-1} n_i n_{i+1} \, , \quad
    H_\mathrm{I}=\sqrt{\frac{\gamma}{\Delta t}} \sum_{i = 1}^L (1-n_i) \otimes \tau_i^x,
    \label{Seq:Rydberg_Hamiltonian}
\end{align}
is a sum of local terms. Since $U_\mathrm{CM}$ is generally nonlocal, it cannot be directly expressed as an MPO with manageable bond dimension. We therefore apply a first order Trotter decomposition,
\begin{align}
U_{\mathrm{CM}} = e^{-i H_{\mathrm{CM}} \Delta t} = \left (e^{-i H_{\mathrm{CM}} \Delta t / N}\right )^N \approx  \left (e^{-i H_{\mathrm{I}} \Delta t / N} e^{-i (H_{\mathrm{S}})_e \Delta t / N} e^{-i (H_{\mathrm{S}})_o \Delta t / N} \right )^N,
\end{align}
where $(H_{\mathrm{S}})_{e(o)}$ is the sum of terms in $H_\mathrm{S}$ acting on even (odd) bonds, using the standard even–odd splitting of nearest-neighbor Hamiltonians. This introduces an error of order $\mathcal{O}((\Delta t^2)/N)$. In our simulations we set $N=10$, which we verified to be sufficient to reproduce exact results for small system sizes.

Fig.~\ref{fig:trotterization} illustrates the MPO representation of this decomposition. The product of exponentials for $H_\mathrm{S}$ terms yields an MPO with bond dimension 2 [Fig.~\ref{fig:trotterization} (a)]. The interaction Hamiltonian $H_{\mathrm{I}}$, being local on each system-ancilla pair, admits a simple product structure [Fig.~\ref{fig:trotterization} (b)], corresponding to bond dimension 1. Their contraction gives the MPO for a single Trotter step [Fig.~\ref{fig:trotterization} (c)], without increasing the bond dimension. A full Kraus operator is then obtained as the contraction of $N$ such Trotter steps [Fig.~\ref{fig:trotterization} (d)]. Applying one collision step $\Delta t$ of the evolution then amounts to $N$ sequential applications of this MPO onto the product of the system state with a fresh chain of ancillas in the product state $\ket{0}_A^{\otimes N}$, and produces the evolved state of system and ancillas.

\subsection{Convergence with bond dimension}
To compute the dominant eigenvalue of the biased transfer matrix defined in~\eqref{eq:biased_map}, we employ a power method adapted to TNs. In each iteration, the MPS is updated by applying the MPO for a full collision step, decomposed into 10 Trotter layers. To avoid exponential bond-dimension growth, truncations are performed after each layer, retaining only a fixed number of singular values. This is achieved using standard singular value decomposition (SVD) at every bond, and discarding singular values below $10^{-14}$, retaining at most a fixed bond dimension $D_\mathrm{max}$. The last step is followed by ancilla tracing, potentially with the bias insertion, so that at the end of the step, the result is a mixed state of the system only described as an MPS (a vectorized MPO).

\begin{figure}[h]
    \centering
    \includegraphics[width=\textwidth]{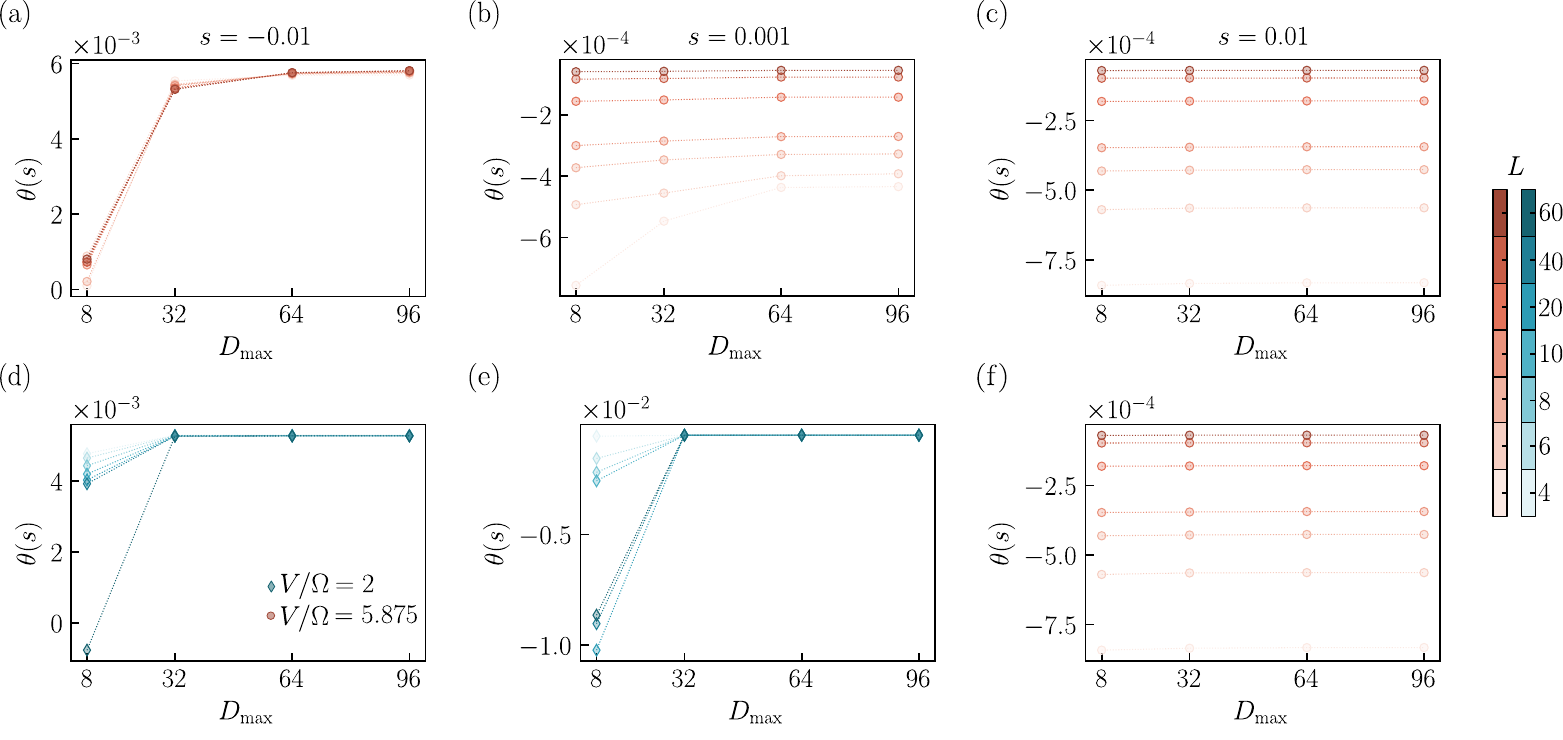}
    \caption{\textbf{Convergence of the power-method results with bond dimension.} Panels show the value of $\theta$ as a function of the maximum bond dimension $D_\mathrm{max}$, for biasing fields $s=-0.01$ (left column), $s=0.001$ (middle column), and $s=0.01$ (right column). The top row corresponds to $V/\Omega=5.875$ and the bottom row to $V/\Omega=2$. In all cases, convergence is reached around $D_\mathrm{max}=96$, as differences with respect to the previous bond dimension become negligible.
    We estimate our errors as the difference between the estimate at the largest bond dimension $D_\mathrm{max}=96$ and a smaller one $D_\mathrm{max}=64$.}
    
    \label{fig:S12}
\end{figure}

The results show clear convergence of observables as a function of $D_\mathrm{max}$, as illustrated in Fig.~\ref{fig:S12}, confirming the reliability of our approach for the chosen bond dimensions.

\section{Exact Computation of the Activity Using an MPO}
Since the activity $a(s)$ is a derivative of the SCGF, $a(s)=-d\theta(s)/ds$, its value can be estimated numerically via finite differences of the computed $\theta(s)$. However, the TN method provides an estimate not only of the dominant eigenvalue $\Lambda(s)=e^{L \theta(s)}$, but also of the corresponding eigenvector approximated as an MPS, which can be used to calculate the activity in terms of an appropriate matrix element.

By definition, the SCGF is related to the leading eigenvalue $\Lambda(s)$ of the tilted channel $\mathcal{E}_s$ via $\theta(s) = \ln \Lambda(s)$ / L. The structure of $\mathcal{E}_s$ (a CP map) guarantees that $\Lambda(s)$ is real, and the corresponding eigenvector $\rho_s$ is a positive operator. Differentiating with respect to $s$ gives the mean activity in the $s$-ensemble,
\begin{align}
    a(s) = -\,\frac{d\theta(s)}{ds} 
    = -\frac{1}{L \Lambda(s)} \frac{d \Lambda(s)}{ds}.
    \label{eq:activity_lambda}
\end{align}

Using $\mathcal{E}_s \ket{\rho_{s}(s)}=\Lambda(s)\ket{\rho_{s}(s)}$, we can write (we drop the explicit dependence on $s$ for simplicity of the notation)
\begin{equation}
\bra{\omega_s}\mathcal{E}_s \ket{\rho_s}=\Lambda(s) 
\end{equation}
where $\bra{\omega_s}$ is the left eigenvector associated with $\Lambda(s)$ that we normalize as $\bra{\omega_s} \rho_s\rangle = 1$. In analogy to the Hellmann-Feynman theorem~\cite{Sakurai_Napolitano_2020}, the derivative of $\Lambda(s)$ then can be expressed as
\begin{equation}
    \frac{d \Lambda(s)}{ds}= \bra{\omega_s}\mathcal{E}_s' \ket{\rho_s}.
\end{equation}
Here, the derivative of the tilted channel is also a CP map,
\begin{align}
    \mathcal{E}_s'=\frac{\partial \mathcal{E}_s}{\partial s} 
    = - \sum_{\mathbf{k}} \Bigg(\sum_{i=1}^{L} {k}_i \Bigg)\,
    e^{-s \sum_{i=1}^{L} {k}_i}\,
    K_{\mathbf{k}} \otimes K_{\mathbf{k}}^\dagger ,
    \label{eq:activity_derivative}
\end{align}
where $\mathbf{k} \in \{0,1\}^L$ labels the binary outcome string and $K_{\mathbf{k}}$ denotes the corresponding Kraus operator. This derivative has the same MPO structure as the tilted channel $\mathcal{E}_s$, but modifying the bias operator in the ancilla legs as
$$
e^{-s \hat{A}} \to -\hat{A} e^{-s \hat{A}}, 
$$
itself an MPO of bond dimension 2. Finally, the activity per site  can be written as
\begin{align}
    a(s) = -\frac{1}{L \Lambda(s)} \bra{\omega_s}\mathcal{E}_s' \ket{\rho_s}.
    \label{eq:act_der}
\end{align}

In our algorithm, we obtain $|\rho_s \rangle$ as an MPS via a power method in which we iterate the application of $\mathcal{E}_s$. Repeating the procedure for the adjoint $\mathcal{E}_s^{\dagger}$, we can obtain the MPS approximation for the corresponding left eigenvector $\langle \omega_s|$. Then the  matrix element in \eqref{eq:act_der} can be computed efficiently as a simple  contraction of an MPO between two MPSs.

Applying this technique requires finding two dominant eigenvectors for each value of $s$. To save computational cost, we have instead used the numerical derivative for most of the range of $s$ values, except at $s=0$, where the numerical derivative is less precise and $\langle \omega_{s=0}|$ is actually known exactly. At this point, $\mathcal{E}_{s=0}$ is a CPTP map, so that its dominant left eigenvector is equal to the identity (maximally mixed state), and Eq.~\ref{eq:act_der} can be computed without running the power method.

\section{Numerical Estimation of the Critical Counting Field \texorpdfstring{$s^*$}{s*}}
We define the critical field $s^*$ as the value of $s$ at which a dynamical phase transition occurs, as observed in the main text. To determine the phase diagram for the dynamical activity shown in Fig.~\ref{fig:phase_diagram}(c), we run simulations for a narrowly spaced grid of interaction values $V/\Omega\in (3,\,8)$, and determine $s^*(V/\Omega)$ for each of them. 

As discussed in the main text, the transition manifests as a sharp crossover between two distinct branches of the SCGF: a linear branch that decreases from negative $s$ and a nearly flat (or weakly curved) branch for positive $s$. The crossing between these two regimes signals the location of the phase transition.

To estimate this crossing point systematically, we proceed as follows. For each value of $V/\Omega$, we first compute the SCGF numerically for a range of positive values of $s$, using the power method. Then, we construct a lower bound for the linear branch using first-order perturbation theory near $s=0$, where $\theta(s) \gtrsim \theta(0) - a_0\cdot s = -a_0\cdot s$, with $a_0$ the mean activity at $s=0$ and $\theta(0)=0$. We identify the two numerical points on the upper (flat) branch closest to this linear bound, while ensuring they lie strictly above it. A linear extrapolation of these two points is then performed, and the crossing point of this extrapolated line with the linear bound is defined as $s^*$.

\begin{figure}[ht]
    \centering
    \includegraphics[width=\textwidth]{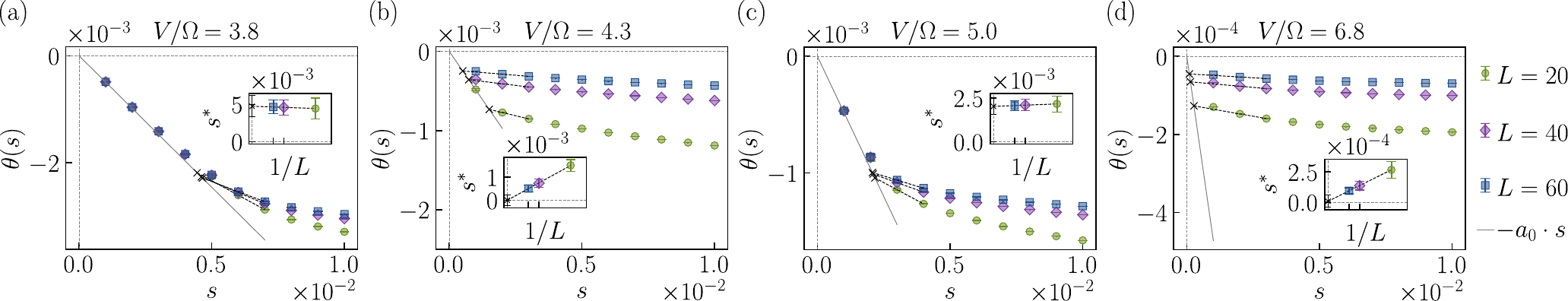}
    \caption{\textbf{Estimation of the critical counting field $s^*$.} 
    SCGF $\theta(s)$ for (a) $V/\Omega=3.8$, (b) $V/\Omega=4.3$, (c) $V/\Omega=5.0$ and (d) $V/\Omega=6.8$. The solid grey line shows the perturbative linear bound $-a_0\cdot s$. Symbols denote numerical results from the power method for different values of $L$.  Insets: linear extrapolation of the two numerical points on the flat branch closest to the bound. The intersection when $1/L\rightarrow 0$ defines $s^*$.}
    \label{fig:S3}
\end{figure}

This procedure is repeated for all values of $V/\Omega$, yielding a consistent set of estimates for $s^*$, which form the transition line in the dynamical phase diagram. Examples of this construction for different $V/\Omega$ are shown in Fig.~\ref{fig:S3}. It is important to note that the uncertainty in the estimate of $s^*$ is larger for values of $V/\Omega$ where the crossover in activity is more gradual and less sharp, indicating a smoother finite-size rounding of the transition.

The resulting numerical estimates of $s^*$ for the interaction strengths shown in
Fig.~\ref{fig:S3} are summarized in Table~\ref{tab:sstar}.

\begin{table}[h]
\centering
\caption{Critical counting field $s^*$ for the values of $V/\Omega$ shown in Fig.~\ref{fig:S3}.}
\label{tab:sstar}
\begin{tabular}{c c}
\hline\hline
$V/\Omega$ & $s^*$ \\
\hline
3.8 & $0.005\pm0.001$ \\
4.3 & $0.0000\pm0.0002$ \\
5.0 & $0.0020\pm0.0004$ \\
6.8 & $0.00001\pm0.00005$ \\
\hline\hline
\end{tabular}
\end{table}


\end{document}